\documentclass[oneside,onecolumn,12pt,a4paper]{article}
\usepackage{graphicx}
\usepackage{dsfont}
\usepackage{natbib}
\usepackage{bbding}
\usepackage{setspace}
\usepackage{braket}
\usepackage{sectsty}
\usepackage[multiple]{footmisc}
\usepackage[margin=20mm]{geometry}
\usepackage{tikz}
\usepackage{bm}
\usepackage{authblk}
\usepackage{amsmath}
\usepackage{lmodern}
\usepackage{commath}
\usepackage{mathtools}
\usepackage{amssymb}
\usepackage{hyperref}
\usepackage{qtree}
\usepackage{color}
\definecolor{darkblue}{rgb}{0.0,0.0,0.3}
\hypersetup{colorlinks,breaklinks,linkcolor=darkblue,urlcolor=darkblue,anchorcolor=darkblue,citecolor=darkblue}

\usetikzlibrary{calc}
\usetikzlibrary{arrows}
\usetikzlibrary{decorations.markings}
\usetikzlibrary{shapes}
\usetikzlibrary{positioning}

\usepackage[justification=justified,textfont=small,width=.9\textwidth]{caption}

\def\citepos#1{\citeauthor{#1}'s (\citeyear{#1})}

\usepackage{ifthen}
\def\eprinttmp@#1arXiv:#2 [#3]#4@{\ifthenelse{\equal{#3}{}}{\href{http://arxiv.org/abs/#1}{arXiv:#1}}{\href{http://arxiv.org/abs/#2}{arXiv:#2 [#3]}}}
\newcommand{\eprint}[1]{\eprinttmp@#1arXiv: []@}
\newcommand{\doi}[1]{\href{http://dx.doi.org/#1}{doi:#1}}

\sectionfont{\normalfont\normalsize\bfseries}
\subsectionfont{\normalfont\small\bfseries}
\subsubsectionfont{\normalfont\small\bfseries}

\makeatletter
\g@addto@macro\quote{\small\singlespacing\upshape\sffamily\vspace{-4mm}}
\makeatother

\begin{document}

\setstretch{1.2}

\title{What can bouncing oil droplets tell us about quantum mechanics?}

\author[1]{\bf Peter W. Evans\thanks{email: \href{mailto:p.evans@uq.edu.au}{p.evans@uq.edu.au}}}
\author[2]{\bf Karim P. Y. Th\'ebault\thanks{email: \href{mailto:karim.thebault@bristol.ac.uk}{karim.thebault@bristol.ac.uk}}}

\affil[1]{\small{{\it School of Historical and Philosophical Inquiry}, University of Queensland}}
\affil[2]{\small{{\it Department of Philosophy}, University of Bristol}}

	\maketitle

\begin{abstract}
	A recent series of experiments have demonstrated that a classical fluid mechanical system, constituted by an oil droplet bouncing on a vibrating fluid surface, can be induced to display a number of behaviours previously considered to be distinctly quantum. To explain this correspondence it has been suggested that the fluid mechanical system provides a single-particle classical model of de Broglie's idiosyncratic `double solution' pilot wave theory of quantum mechanics. In this paper we assess the epistemic function of the bouncing oil droplet experiments in relation to quantum mechanics. We find that the bouncing oil droplets are best conceived as an \textit{analogue illustration} of quantum phenomena, rather than an \textit{analogue simulation}, and, furthermore, that their epistemic value should be understood in terms of how-possibly explanation, rather than confirmation. Analogue illustration, unlike analogue simulation, is not a form of `material surrogacy', in which source empirical phenomena in a system of one kind can be understood as `standing in for' target phenomena in a system of another kind. Rather, analogue illustration leverages a correspondence between certain empirical phenomena displayed by a source system and aspects of the ontology of a target system. On the one hand, this limits the potential inferential power of analogue illustrations, but, on the other, it widens their potential inferential scope. In particular, through analogue illustration we can learn, in the sense of gaining how-possibly understanding, about the putative ontology of a target system via an experiment. As such, the potential scientific value of these extraordinary experiments is undoubtedly a significant one.
\end{abstract}

\newpage
\tableofcontents

\section{Introduction}
\label{sec:intro}

In 2005, a team in Paris Diderot University led by Yves Couder and Emmanuel Fort discovered that an oil droplet bouncing on a vibrating fluid surface can be made to `walk' horizontally across the surface. These `walkers' display a kind of wave-particle duality: the bouncing droplet is self-propelled by interacting with the surface waves it creates. A series of subsequent experiments from both the team in Paris and an associated team led by John Bush at Massachusetts Institute of Technology (henceforth, the `walker experiments') have since demonstrated that this fluid mechanical system displays behaviour that is typically considered to be quantum behaviour. This behaviour includes single and double slit diffraction and interference \citep{Couder06}\footnote{Although recent experiments contest the single and double slit diffraction and interference results: at best these phenomena are difficult to reproduce \citep{Pucci2018}, and at worst cannot be reproduced at all \citep{Andersen2015,Batelaan_2016,Bohr2016}.} and quantised orbits of bound state pairs \citep{Fort10}, as well as phenomena that look analogous to quantum tunnelling \citep{Eddi09}, Schr\"{o}dinger evolution of probabilities \citep{Couder12}, and Zeeman splitting \citep{Eddi12}.

A string of strongly qualified suggestions have emanated from both the Paris and MIT teams that this fluid mechanical system provides a single-particle classical model of the pilot wave mechanism of de Broglie's idiosyncratic double solution pilot wave theory. For instance (emphasis added): ``Such a system is \textit{reminiscent} of the early de Broglie models for quantum systems'' \citep[p.2]{Couder12}; ``This \textit{appears very close} to the hypothesis of a double solution put forward by de Broglie'' \citep[p.6]{Couder12}; ``our system \textit{could be considered} as implementing at [a] macroscopic scale the idea of a pilot wave considered by de Broglie and Bohm for elementary particles at [a] quantum scale'' \citep[p.17520]{Fort10}; ``in spite of the huge gap between the systems, \textit{analogies exist with some aspects of} the particle-wave behaviour at [a] quantum scale'' \citep[p.459]{Eddi11}; ``This hydrodynamic system \textit{bears a remarkable similarity} to an early model of quantum dynamics, the pilot wave theory of Louis de Broglie'' \citep[p.613]{molacekbush2013}; ``The walker system \textit{bears a notable resemblance} to an early conception of relativistic quantum dynamics, Louis de Broglie's double-solution pilot-wave theory'' \citep[p.170]{Bush15}.

The basic idea behind this suggestive correspondence is that the motion of the oil droplet (the `walker') is determined by the wave on the fluid surface in just the same way that the motion of a particle is determined by the phase of an associated wave in de Broglie's double solution formulation of the pilot wave approach to quantum theory. It is interesting to ask, however, whether and to what extent this classical fluid mechanical system can be considered a genuine \textit{epistemic tool} to probe quantum behaviour. Our analysis is framed by a two-way comparison between these experiments and recent philosophical discussions of confirmation via analogue black hole experiments \citep{Dardashti2015} and explanations via toy models \citep{Reutlinger17}. We propose that, despite a superficial similarity, the epistemic function of the walker experiments is very different to that of analogue black hole experiments.

The analogue black hole experiments, as reconstructed by \citet{Dardashti2015}, exemplify the increasingly common scientific practice of analogue simulation, where a material system is manipulated and a formal relationship obtains between \textit{empirical terms} (i.e. linguistic items that putatively correspond to physical phenomena) in the model of the system being manipulated (the source) and the model of the system about which we hope to learn (the target). The source system is taken to be a `material surrogate' -- to `stand in for' -- the target system. In such circumstances it is possible that we can gain knowledge about \textit{actual} phenomenal features of the target system.

In contrast, in the case of the walker experiments, according to our rational reconstruction, the material system is manipulated with the epistemic goal of establishing a formal relationship between empirical terms in the source model (i.e. the model of the walker) and \textit{extra}-empirical terms in the target model (i.e. terms corresponding to non-phenomenal aspects of the ontology of de Broglie's double solution theory). We categorise such activities as \textit{analogue illustration}, and argue that we should understand their primary epistemic function as the provision of \textit{how-possibly explanations}. In this precise sense, our account of the epistemic function of the walker experiments has close parallels to the account of explanations via toy models provided by \citet{Reutlinger17}.

Analogue illustration, unlike analogue simulation, is not a form of `material surrogacy' in which source empirical phenomena in a system of one kind can be understood as `standing in for' target phenomena in a system of another kind. Rather, analogue illustration leverages a correspondence between certain phenomena displayed by a source system and aspects of the ontology of a target system. On the one hand, this limits the potential inferential power of analogue illustrations, but, on the other, it widens their potential inferential scope. In particular, through analogue illustration we can learn, in the sense of gaining how-possibly understanding, about the putative ontology of a target system in an entirely novel way. As such, the potential scientific value of these extraordinary experiments is undoubtedly a significant one.

We proceed as follows. In \S\ref{sec:pilot} we describe in detail both de Broglie's double solution pilot wave theory and the walker experiments, with a view to identifying the relevant connections between the two. In \S\ref{sec:analogy} we discuss the various relevant ideas that can be drawn from the philosophical analysis of analogue experiments in general. First, we introduce the crucial distinction between analogue illustration and analogue simulation. Second, we consider the subtle issue of the validation of illustrations and simulations, and the question of whether a single experiment can combine both functions. Finally, we employ these distinctions to examine the epistemic value of different forms of analogue experiments with a focus on confirmation and explanation. \S\ref{sec:Parisanalogue} then deploys this general framework to provide an answer to the question: What can bouncing oil droplets tell us about quantum mechanics? We consider in turn the simulation, illustration and explanatory aspects of the experiments and offer our constructive, if rather deflationary, conclusion: we find that the bouncing oil droplets are best conceived as \textit{analogue illustrations} of certain quantum phenomena, rather than \textit{analogue simulations}, and that their epistemic value should be understood in terms of how-possibly explanation, rather than confirmation. In addition we warn that, due to the classicality of the walker experiments, their ability to be an analogue illustration (let alone simulation) of any entanglement-based quantum phenomena that involve violation of Bell-type inequalities is exceedingly constrained.

\section{De Broglie's pilot wave theory and the walker experiments}
\label{sec:pilot}

Pilot wave theory was first proposed by \citet{deBroglie24,deBroglie27a,deBroglie27b} before being independently redeveloped twenty five years later by \citet{Bohm}. The formulation emphasises the existence of an actual configuration of particles that underpins the dynamics of a quantum system, where the actual positions and velocities of the constituent particles comprise a set of `hidden variables' for the system. The claims concerning analogy emanating from the Paris and MIT teams mostly concern specifically de Broglie's pilot wave theory. It will prove worthwhile to briefly set out the basic distinction between de Broglie's original formulation of pilot wave theory and Bohm's subsequent formulation before we provide a more detailed account.

For de Broglie, quantum dynamics is characterised by the Schr\"{o}dinger equation, which defines a `pilot wave' that governs the evolution of the configuration of the system on configuration space, and a guiding equation, which determines the velocities of the particle trajectories prescribed by the evolution of the configuration as a function of a phase defined by the pilot wave. This velocity guiding equation is taken by de Broglie as the fundamental law of motion and provides a first-order dynamics in his formulation of pilot wave theory \citep[p.29]{BacciagaluppiValentini}. For Bohm, quantum dynamics is characterised as an extension of classical dynamics with the addition of a `quantum potential' (first derived by de Broglie) that is a function of the solution to the Schr\"{o}dinger equation, and which contributes a quantum force on the particle configuration in addition to any classical Newtonian forces. The Newtonian relation between the quantum potential and the acceleration of the particle trajectories is taken by Bohm as the fundamental law of motion and provides a second-order dynamics in his formulation of pilot wave theory, while de Broglie's fundamental guiding equation is taken in Bohm's formulation as a dynamical constraint \citep[p.29]{BacciagaluppiValentini}. Thus the formal apparatus of both de Broglie's and Bohm's formulations of pilot wave theory are identical, the difference between the two approaches is whether the quantum description is \textit{fundamentally} Newtonian or not -- Bohm adopting the former and de Broglie adopting the latter. This crucial difference between the two approaches thus amounts to a difference of \textit{extra-empirical} structure. It will be worthwhile for the reader to keep this basic point of difference in mind as we dive into the detailed description that follows.

\subsection{Double solution theory}

Undoubtedly de Broglie's most famous contribution to quantum theory is the extension of \citepos{Einstein1905} idea of wave-particle duality from photons to massive particles. This idea is first articulated in de Broglie's doctoral thesis \citep{deBroglie24} but is not the central focus of that work. Rather, de Broglie's aim was to propose an equivalence between Fermat's principle of least time for describing rays of light and Maupertuis' principle of least action for describing moving bodies \citep[p.56]{deBroglie24}. In particular, the equivalence of Maupertuis' and Fermat's principles suggests that the role played by the 4-momentum, $p_{\mu}$, in the motion of a body corresponds to the role played by the phase differential, $\dif \phi$, in the propagation of a wave. Equating these two elements amounts essentially to a 4-vector generalisation of the Planck-Einstein relation, $E = h \nu$, to $p_{\mu} = h w_{\mu}$, with $w_{\mu} \propto \dif \phi$ \citep[p.42]{BacciagaluppiValentini}.

This generalisation amounts to a new law of motion for de Broglie -- an early statement of his guiding equation -- in which the momentum of a body (or particle velocity) is determined by the phase of an associated wave.\footnote{For a detailed exploration of the connection between this aspect of de Broglie's work and Schr\"{o}dinger's derivation of the equation that bears his name see \citep{joas:2009}.} Significantly, such a law of motion is a departure from Newtonian mechanics: the law of inertia no longer applies to these quantum bodies, which would move along the rays of their associated wave, and so can deviate from a straight path without any applied forces (for instance, in the process of diffraction).\footnote{\Citet[p.80]{deBroglie24} does presciently note, however, that one could recover the Newtonian picture by imagining a force to be active in such a process. Of course, this force is a function of the `quantum potential' which Bohm was later to emphasise.}

The full presentation of de Broglie's theory -- the one that we have come to know as pilot wave theory -- is given in \citep{deBroglie27a}. In contrast to Schr\"{o}dinger, de Broglie explicitly hypothesises that particles be understood as solutions of the wave equations of motion, ``the amplitude of which includes a peculiar singularity'' \citep[p.225]{deBroglie27a}, such that this `singular amplitude' is to be understood as a single particle. Whereas Schr\"{o}dinger develops a perspective in which `particles' are an unnecessary part of the dynamics of the continuous phase wave $\Psi$ (and Born goes one step further and explicitly attributes a statistical nature to this wave), \citet[p.225]{deBroglie27a} is motivated by the possibility of a duality between Schr\"{o}dinger's continuous solutions of the wave equations and his singular solutions representing the particles (hence his `double solution' theory). He begins with the relativistic wave mechanical equation of motion (the Klein-Gordon equation) as a description of a particle in a constant potential, and considers solutions, $u(\mathbf{x},t)$, with singular amplitude (that is, solutions representing a single particle). From this he is able to derive a partial differential equation which, in the regime in which the behaviour of the amplitude singularity obeys the classical wave equation, approximates the relativistic equation of motion for a classical particle. Since in this classical regime the particle velocity aligns with the phase gradient, \citet[p.230]{deBroglie27a} assumes that it does so in the nonclassical regime also, reinforcing his strong commitment to the equivalence of the principles of Maupertuis and Fermat (particle and wave descriptions).

De Broglie then considers a continuous solution, $\Psi(\mathbf{x},t)$, to the same wave equation (which is identical to the solutions to Schr\"{o}dinger's wave equation). Whereas, in modern parlance, the $u(\mathbf{x},t)$ are to be interpreted as ontic, since they represent real physical particles, the $\Psi(\mathbf{x},t)$ are interpreted as epistemic, since they are taken to represent an \textit{effective ensemble} of the singular particles. De Broglie derives a partial differential equation which, in the regime in which the continuous amplitude, $a$, is harmonic, i.e., $\nabla^{2}a(\mathbf{x})=0$, approximates the wave equation for geometrical optics. He notes that the relativistic equation of motion for a classical particle and the wave equation for geometrical optics are the same equation of motion (the Hamilton-Jacobi equation) so long as the phase factors in each are identical. He then assumes that the phase factors are identical in the general (nonclassical) case, which ensures that the singular and continuous wave solutions are interdependent by virtue of sharing a common phase factor. It is this assumption that de Broglie refers to as the `principle of the double solution', and this interdependence between the singular and continuous waves is the embodiment of the wave-particle duality at the core of de Broglie's double solution theory.

\Citet[p.232]{deBroglie27a} demonstrates that, from the identity of phase factors, one can understand $\Psi$ as representing an effective ensemble particle density over space, where density is proportional to the square of the amplitude of $\Psi$. Moreover, de Broglie realises that interpreting $\Psi$ as a representation of particle density has a novel consequence: equal probability over the initial positions of an ensemble of particles results in a set of possible particle trajectories and, since the trajectories are determined by surfaces of equal phase, the trajectories are \emph{guided by} $\Psi$ (\citealp[p.232]{deBroglie27a}; \citealp[p.59]{BacciagaluppiValentini}). For this reason, $\Psi$ is called the guiding wave, or pilot wave, and the generalised quantum relation relating the phase of the guiding wave to the particle velocity is known as the guiding equation.

\subsection{Ontology of the double solution theory}
\label{subsec:understand}

To understand the ontology of de Broglie's double solution theory it will prove highly instructive to contrast his interpretation with those of Sch\"{o}dinger and Bohm respectively.

Schr\"{o}dinger was adamant that the continuous phase wave $\Psi$ is not accompanied by an associated particle with a well-defined position or trajectory, rather `particles' are to be identified with the spatially distributed wave packet \citep[p.116]{BacciagaluppiValentini}. However, \citet[p.238]{deBroglie27a} notes, when one considers a many-body system of $N$ particles, the solutions to the relevant wave equation exist in $3N$-dimensional configuration space. If this were the case, then it would be difficult to see the physical meaning of the coordinates used to construct the abstract configuration space in the first place. Moreover, it is unclear what physical meaning we should attach to the propagation of the continuous wave in an abstract configuration space. De Broglie points out that both of these difficulties disappear if we admit that particles always have well-defined positions and trajectories and that the physical picture of the system consists in $N$ waves propagating in real 3-space rather than a single wave propagating in $3N$-dimensional configuration space.

This then is the foundation of de Broglie's interpretation of the pilot wave. By extending his framework for understanding $u(\mathbf{x},t)$ and $\Psi(\mathbf{x},t)$, and the phase identity between them, from the single-particle case to the many-body case (where there would now be many $u_{i}(\mathbf{x},t)$), de Broglie is able to interpret $\Psi$ as performing two distinct roles. Firstly, $\Psi$ is a pilot wave: its phase, by being identical to the phase of the singular waves, determines the particle velocities of the system (and so the particle trajectories when the initial positions are given). Secondly, $\Psi$ is a probability wave: the ``fictitious'' wave in configuration space in the many-body case plays the same role as the continuous wave in the single-particle case such that the square of its amplitude determines at each point the ``probability of presence'' of the particle configuration (and so the probability density of the particle trajectories when the initial positions are not given) \citep[p.240]{deBroglie27a}.

What we must not lose sight of here is that, for de Broglie, the guiding equation is the fundamental equation of motion for his new dynamics. That is, the insight at the heart of the equivalence between the principles of Maupertuis and Fermat -- that there is phase harmony between the wave and particle aspects of the double solution to the dynamical equation -- is the key motivation for de Broglie's pilot wave theory. So while $\Psi$ plays the role of the pilot wave, it can only do so as a result of this phase harmony between it and the real $u$-waves in 3-space. The reason it is important to mention this here relates to a possibility that \citet[p.241]{deBroglie27a} points out for understanding $\Psi$. De Broglie's velocity law of motion is derived by invoking his double solution principle. Recall that as a result of the principle, de Broglie simply assumes that the phase factors in the particle and wave motions are identical in the general, non-classical case on account of establishing that the particle velocity aligns with the phase gradient in the classical case. But one could just as easily assume the velocity law to hold as a \emph{postulate} of the theory, rather than derive it from some underlying foundation. As a consequence, one could consider the pilot wave $\Psi$ as physically real, and distinct from the reality of the material point, such that the motion of the material point is determined as a function of the phase of the pilot wave by the velocity law \citep[p.64]{BacciagaluppiValentini}, just as Bohm later suggested.

De Broglie does not advocate this position as the appropriate way to understand his theory. Indeed, \citet[p.241]{deBroglie27a} is explicit about considering this understanding of $\Psi$ as a real pilot wave in this sense with a ``provisional attitude'', preferring the above account of his double solution theory -- and believing that the provisional account would ultimately lead one towards something analogous to the double solution theory in any case. However, \citet{deBroglie27b} went on to present precisely this provisional theory at the fifth Solvay conference in 1927, with no use made of the principle of the double solution \citep[p.67--69]{BacciagaluppiValentini}. It was this provisional theory that became known as de Broglie's pilot wave theory, and it was this theory that Bohm rediscovered 25 years later, rectifying some of the outstanding problems identified with de Broglie's presentation of the theory at the fifth Solvay conference. Thus, in so far as Bohm's pilot wave theory is operationally equivalent to quantum mechanics \citep[p.166]{Bohm}, so too is de Broglie's truncated `provisional' theory operationally equivalent to quantum mechanics. But this equivalence belies the significant foundational (interpretational) differences between de Broglie's original double solution formulation and Bohm's later formulation.

The major divergence is that de Broglie takes his guiding equation to be a novel, non-Newtonian, fundamental law of motion. De Broglie is well aware of the possibility that the addition of a Newtonian force, underpinned by a new quantum potential, could plausibly account for the noninertial motion that results from this law, and he explicitly derives this classical formulation of the particle dynamics; in the nonrelativistic limit, the additional potential energy term is precisely Bohm's `quantum potential' (\citealp[p.237]{deBroglie27a}; \citealp[p.61]{BacciagaluppiValentini}). However, for de Broglie this is merely a demonstration of the relation between his new mechanics and classical mechanics. In contrast, the Newtonian relation between the quantum potential and the acceleration of the particle trajectories is taken by Bohm to be the fundamental law of motion, and this provides a second-order dynamics (while de Broglie's guiding equation is taken in Bohm's formulation as a dynamical constraint).

The divergence between de Broglie and Bohm is thus manifest at the \textit{ontological} level. Whereas for \citet[p.170]{Bohm} the pilot wave $\Psi$ is ``a mathematical representation of an objectively real field'', for de Broglie $\Psi$ is a `fictitious' probability wave and it is the $u_{i}$ that are real waves in 3-space. Despite the fact that de Broglie never completed his double solution theory -- notwithstanding later attempts \citep{deBroglie59,deBroglie71}) -- de Broglie had a clear idea of the ontology of his theory: an $N$-body system consists of $N$ singular $u$-waves propagating in real 3-space, each defining particle motion according to the velocity equation, and whose phase is interrelated with the phase of the probability wave $\Psi$ that guides and constrains system behaviour. Understanding the walker experiments as a concrete \textit{illustration} of this ontology will prove crucial to our analysis of the epistemic function of the experiments in the remainder of the paper.

\subsection{The walker experiments}
\label{sec:Parisexp}

\begin{figure}[t]\vspace{8mm}
  \begin{center}
    \begin{tikzpicture}[scale=0.5]
      \draw (-0.75,-1.5)--(-0.75,1.5); 
      \draw (0.75,-1.5)--(0.75,1.5); 
      \draw [stealth-stealth] (0,-0.75)--(0,0.75); 
      \draw (-0.75,0)--(-5,0); 
      \draw (0.75,0)--(5,0); 
      \draw (-5,0)--(-5,-3); 
      \draw (5,0)--(5,-3); 
      \draw (-2.5,-1.5) rectangle (2.5,-2); 
      \filldraw [fill=blue, draw=blue] (-10,1.5) rectangle (10,2.25); 
      \draw (-10,1.5)--(10,1.5); 
      \draw (-10,1.5)--(-10,3.5); 
      \draw (10,1.5)--(10,3.5); 
      \draw [stealth-stealth] (10.5,1.75)--(10.5,3.25); 
      \filldraw [fill=blue, draw=blue] (-3,2.8) circle (1.5mm); 
    \end{tikzpicture}
  \end{center}
  \vspace{3mm}
  \caption{The basic setup of the walker experiments. The bath is driven by a low frequency generator to vibrate vertically, sustaining the oil droplet to bounce on the fluid surface indefinitely. The externally driven vertical vibration also sustains the Faraday standing waves generated on the surface of the fluid by the bouncing of the droplet.\label{fig:bath}}
\end{figure}
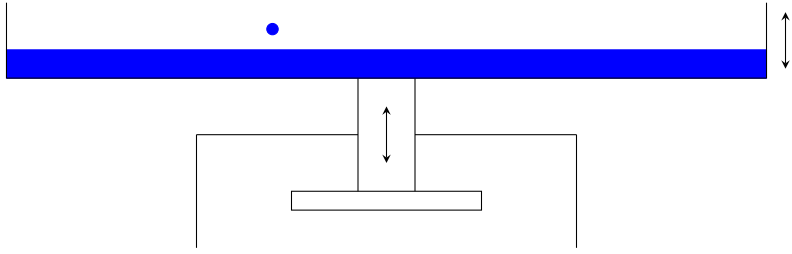

Consider a small, shallow rectangular bath oriented horizontally, filled with a layer of silicon oil, and parametrically driven from below by a low frequency generator to vibrate vertically. By piercing the fluid surface with a pin and then withdrawing quickly, a small oil droplet can be created which, due to the forced vibrations, bounces upon the fluid surface. Fig.~\ref{fig:bath} shows this basic setup of the experiment.

\begin{figure}[t]\vspace{3mm}
  \begin{center}
    \includegraphics[height=0.2\textheight]{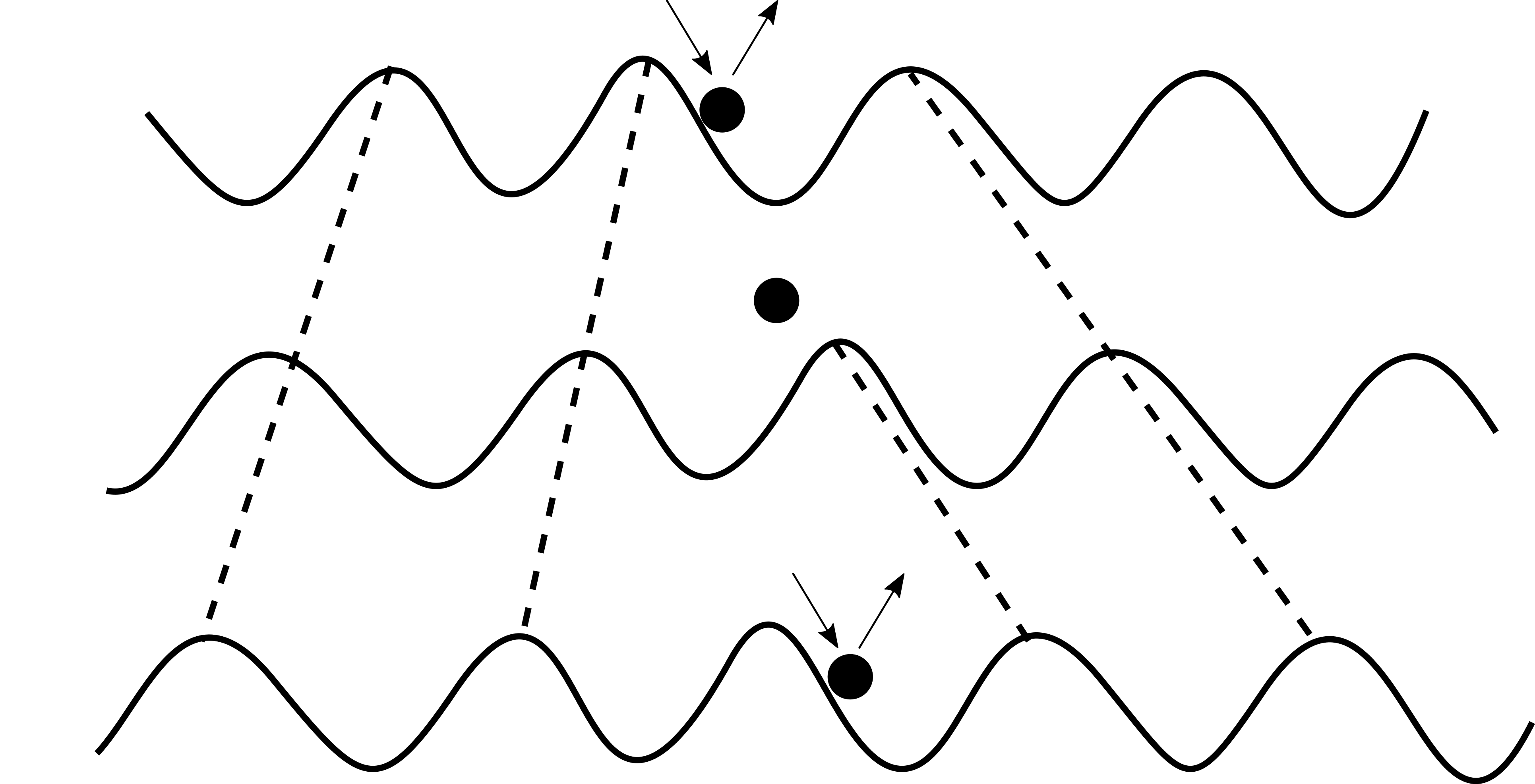}
  \end{center}
  \vspace{3mm}
  \caption{As the droplet is in flight above the surface, the surface wave evolves freely. Small deviations in the flight of the droplet can cause the impact of the bounce to occur on an inclined surface to one side of the central crest of the capillary wave. The gradient of the surface wave imparts a horizontal momentum to the droplet such that the next surface impact again occurs on an inclined surface, sustaining the horizontal motion. (Adapted from \citep[Fig.~6,~p.~93]{Protiere06}.)\label{fig:bounce}}
\end{figure}

There are two key features of this experiment. When the external vibration is strictly below the Faraday instability threshold,\footnote{The phenomenon of Faraday instability is closely related to the more familiar phenomenon of grains of sand on the surface of a beaten drum forming geometrical patterns \citep{Faraday1831}.} beyond which forced standing waves spontaneously appear on the surface of the fluid, the fluid surface is stable in the absence of an oil droplet. When a bouncing droplet is created, and then comes into contact with the surface, it emits a travelling capillary wave across the surface (like a pebble dropped into a still pond) radially damped by the viscosity of the fluid. Call this the `surface wave'. The first key feature of the experiments is that any such droplet created to bounce on the fluid surface is a local source of a standing Faraday wave -- the surface wave -- that is sustained across the fluid surface by the externally driven, vertically vibrating fluid \citep{Protiere05}. Thus, while the droplet is the source of the particular surface wave, the externally driven vertical vibration is what sustains the wave form. During the time that the droplet is in flight above the fluid, this capillary wave evolves freely across the surface. When the droplet next bounces on the surface, small deviations in the flight of the droplet can cause the impact to occur to one side of the central crest of the capillary wave, and thus on an inclined surface. This deviation imparts a horizontal momentum to the droplet such that the next surface impact again occurs on an inclined surface (see Fig.~\ref{fig:bounce}). As \citet[p.~92]{Protiere06} put it:
\begin{quote}
  Each time the drop hits the surface a new dip forms, shifted from the trough that would have been formed by the evolution of the previous wave-packet. The resulting wave is thus the superposition of waves generated by a source that is slightly displaced at each jump.
\end{quote}
Within the appropriate parameter regime, the Faraday wave spontaneously propels the droplet to `walk' horizontally across the surface, coupling the motion of the droplet to the vertical displacement of the fluid surface \citep{Protiere06}. Call this bouncing oil droplet in motion across the fluid surface the `walker'.

Significantly, the damping of a capillary wave generated by a bouncing droplet is inversely proportional to the distance to the Faraday instability threshold: the closer the driving frequency gets to the threshold, the longer the damping length becomes. This is then the second key feature of the walker experiments: due to the external stimulus applied to vibrate the fluid, the standing Faraday waves created from each bounce of the walker are sustained (again, in the appropriate parameter regime) for very many bounces of the walker. The vertical displacement of the surface wave at any point is thus the linear combination of very many distinct Faraday waves, and this provides the surface wave with a path memory of where the walker has recently been \citep{Eddi11}. In other words, the increase in damping length gained by driving the vibration closer to the Faraday instability threshold enables a greater radius of interaction between the walker and its environment.

The combination of the two key features of the walker experiments -- that each droplet is a local source of a standing Faraday wave, and that the standing Faraday waves from each bounce of the droplet are sustained for very many bounces -- yields a surface wave that is a superposition of many standing waves, each encoding information about the previous movement of the walker. The subsequent motion of the walker, bouncing off the fluid surface, is then a function of the information encoded in the surface wave \citep[p.461]{Eddi11}:
\begin{quote}
  Here, each bounce of the droplet, by disturbing the interface, records information about the spatio-temporal localization of the collision. This information is stored because each bounce generates a sustained localized state of Faraday waves. The information being stored in waves, the data about the trajectory are cumulated in an interference pattern due to the waves' linear superposition. Later, as the drop collides again with the interface, it `reads' this cumulated information and the local slope of the distorted surface determines the direction and amplitude of the next jump. The dual nature of the walker is contained in the path memory dynamics: the wave nature lies in the coding while the particle nature lies in the reading.
\end{quote}
This is how the walker comes to be coupled to the surface wave, and so provide what we might call pilot wave dynamics: the ``interplay between the droplet motion and its associated wave field makes it a macroscopic implementation of a pilot wave dynamics'' \citep[p.2]{Couder12}.

Another feature of the walker experiments, related to path memory, has received far less attention (although \citet{Vervoort16} gestures towards it). Each time the droplet bounces on the fluid surface the distinct damped travelling capillary wave that is emitted\footnote{``These waves are\ldots the travelling equivalent of the standing Faraday waves usually observed'' \citep[p.~91]{Protiere06}.} travels at a velocity typically about 10 times the walking velocity of the particle \citep[p.~95]{Protiere06}. Given a boundary -- such as a wall, a slit, a submerged barrier or a new particle with its own associated wave -- within the damping length of the capillary wave, the surface wave that results from the superposition of successive capillary waves will contain a reflected component encoding information about this boundary. Since the particle is coupled to this wave, the motion of the particle is in part influenced by spatially remote boundaries (so long as they are within the damping length), producing what we might call `nonlocal' effects. Significantly, some of these nonlocal boundaries lie in the `future' path of the walking particle and it is this ``echo-location'' feature that explains, for instance, the diffraction of a particle walking through a single slit \citep[p.2]{Couder12}.

A combination of the path memory and these `nonlocal' effects provides for some interesting properties of the walker's motion, the most significant being chaotic motion in the presence of obstacles. As \citet[p.461]{Eddi11} state:
\begin{quote}
  Other dramatic effects of the memory are observed whenever boundaries generate any kind of confinement of the walker. In these situations, the waves emitted in the past and reflected by the boundaries lead to a complex structure of the interference field and correspondingly to a disorder in the droplet motion.
\end{quote}

In a more recent set of experiments performed by the team at MIT, in collaboration with the Paris team, a further significant phenomenon has been detected \citep{Harris2013}. Given a defined bath geometry, a walker will move chaotically under the guidance of the surface wave. After enough time, taking a record of the walker's total path yields a location density map representing the statistical behaviour of the walker. This map can be interpreted as the probability distribution for the walker's location at some time (it turns out that this probability distribution is related to the Faraday modes of the bath geometry). Call this map the walker's probability wave. There are then two wavelike modes of description for the walker's behaviour: a surface wave (that is a superposition of capillary waves) guides (chaotically) the motion of the walker, and a probability wave (that is a function of the geometry of the fluid boundary) constrains the distribution of the locations at which the walker might be found at any time. Here is \citet[p.011001-4]{Harris2013} on this point:
\begin{quote}
  We can thus understand the probability distribution as being a manifestation of the characteristics of the underlying trajectories. In the confined circular geometry, the pilot wave dynamics tends to drive the walker along circular orbits with radii corresponding to maxima of the cavity mode amplitude. Instead of being trapped on these orbits as in the low-path-memory limit, the walker wobbles around them and drifts between them; nevertheless, these unstable orbits leave their mark on the probability distribution.
\end{quote}

This then completes the collection of phenomena that are the significant features of the walker experiments which suggest an analogy with de Broglie's double solution theory. Echoing our sentiment at the end of \S\ref{subsec:understand} above, \citet[p.011001-4]{Harris2013} summarise nicely:
\begin{quote}
  Our study indicates that this hydrodynamic system is closely related to the physical picture of quantum dynamics envisaged by de Broglie, in which rapid oscillations originating in the particle give rise to a guiding wave field. The pilot wave theories of de Broglie and Bohm are often conflated; however, it is valuable to distinguish between them here for the sake of comparison with our system. According to Bohm, the particle is guided by its statistical wave, its velocity being equivalent to the quantum velocity of probability. According to de Broglie's double-wave solution, the particle is guided by a real wave (of unspecified origins) in such a way as to execute a dynamics whose statistics is described by standard quantum theory.
\end{quote}

It should be clear that there are some striking analogies between the mechanism of the walker experiments and the ontology of de Broglie's double solution theory.\footnote{This point is also made by \citet{Bush15}.} At the heart of these analogies is wave-particle duality. In the walker experiments there is a clear interdependency between the motion of the walker and the surface wave of the fluid: the bouncing of the walker generates the surface wave, and the form of the surface wave guides the imminent trajectory of the walker. This interdependency between the surface wave and the walker appears analogous to the interdependency in de Broglie's pilot wave theory between each $u$-wave in 3-space and its associated singular amplitude. Through de Broglie's guiding equation, the particle-like behaviour of these singularities is closely aligned with the wave-like behaviour of their associated $u$-wave. Likewise, the particle-like behaviour of the walker is closely aligned with the wave-like behaviour of the surface wave.

Furthermore, there is an apparent analogy between the walker's probability wave and the pilot wave $\Psi$ in de Broglie's theory. The walker's probability wave represents the statistical behaviour of the walker and constrains the distribution of the locations at which the walker might be found at any time, and $\Psi$, living in configuration space, guides system behaviour through constraining probability current density. Significantly, the differentiation between the walker's probability wave and the surface wave of the fluid superficially resembles the differentiation in de Broglie's double solution theory between $\Psi$ and the real $u_{i}$ in 3-space determining the behaviour of quanta via the guiding equation. Recall that the interconnection between $\Psi$ and $u_{i}$ for de Broglie is formalised in the double solution principle -- they share identical phase factors -- and forms the central pillar of de Broglie's idiosyncratic double solution formulation of pilot wave theory. It is primarily for this reason that the fluid mechanical walker system more closely resembles the ontology of de Broglie's double solution theory than that of Bohm's pilot wave theory. Finally, we should not forget in the context of this apparent or superficial analogy the lengthy list of typically quantum phenomena displayed by the fluid mechanical system: (possibly) single and double slit diffraction and interference, quantised orbits of bound state pairs, phenomena that look like quantum tunnelling, Schr\"{o}dinger evolution of probabilities, and Zeeman splitting (but with the conspicuous absence of any entanglement-based quantum phenomena that involve the violation of Bell-type inequalities).

It remains to be seen whether this analogy implies that these experiments can tell us something about quantum mechanics. What value do these experiments have as epistemic tools to probe quantum theory? In order to answer this question, it will be helpful to consider a selection of issues relating to the philosophical analysis of analogue experiments in general terms.

\section{Analogue experiments}
\label{sec:analogy}

\subsection{Simulation and illustration}

Before we start our discussion it will be instructive to introduce some terminology relating to different parts of scientific theories and models and their putative correlates in the world. When we use the word \textit{model} without further qualification we will be indicating a set of equations together with modelling assumptions and relevant theoretical concepts and (sometimes approximate) laws. Examples of models in this sense are models of continuum fluid dynamics, models of motion in a central gravitational potential, and models of simple harmonic motion. Sometimes for clarity we will refer to this sense of model as an \textit{abstract model}. This is in contrast to the other relevant sense of model, which is as a concrete object used in scientific practice. Examples of models in this sense are plastic molecular models, miniature models of planes used in wind tunnels, and clock-work orrery models of the solar system. We will always refer to models in this sense as \textit{concrete models}. Next, we will use the word \textit{term} to denote any linguistic items within a theory or (abstract) model. This includes terms in the logical sense, statements, equations, diagrams, or structures. Such `terms' putatively stand in a relation of \textit{representation} to elements of reality within the world. The precise nature of this representation relation is something regarding which we will endeavour to stay as neutral as possible -- we take what we say below to be compatible with any of the various accounts in the contemporary literature.\footnote{An excellent recent discussion specifically relating to representation via material models is \citep{frigg:2018}. Further accounts, all of which we take to be compatible with our use of `representation' below, are \citep{hughes:1997,giere:1999,suarez:2004,contessa:2007,BailerJones2009,weisberg:2012}. A good overview of various connected issues is provided in \cite[\S2]{gelfert:2016}.\label{fn:repres}} We can distinguish three different classes of terms on the basis of the three different classes of elements of reality to which they can putatively correspond. First, we have \textit{observable terms} that correspond to observable phenomena. In the context of physical science such observable phenomena will typically be physical quantities whose value can be directly measured or observed -- for example, the angular diameter of the Moon as measured from the Earth. It is important to distinguish between these terms and \textit{empirical terms} that  correspond to both observable and unobservable phenomena. That is, a larger set of elements of reality that (in the context of physical science) also typically includes physical quantities the value of which cannot be directly measured or observed, but rather only indirectly measured or inferred. The most vivid example of such a term is probably the mass of the Higgs boson, but various other examples can be found in both historical and contemporary physics.   

As powerfully argued by \cite{massimi:2007}, building on the earlier ideas of \cite{bogen:1988}, such unobservable phenomena are what is typically `saved' by scientists in experimental particle physics, and should thus be taken as part of any adequate empiricist philosophy of science, \textit{pace} \cite{van:1980}. Further discussion of the observable versus unobservable phenomena distinction is found in \cite{Evans:2020}. For the purpose of the current paper, the more important distinction is between empirical terms and what we will call \textit{extra-empirical terms}. These are linguistic items within a theory or model that putatively stand in a relation of representation to elements of reality within the world that are \textit{non-phenomenal}. That is, elements of reality that are \textit{not} physical quantities. The most powerful examples of such non-phenomenal elements of reality, to which we will refer back later, are absolute space in Newtonian mechanics and the wavefunction in quantum mechanics.      
     
\begin{figure}
  \centering
  \includegraphics[height=0.3\textheight]{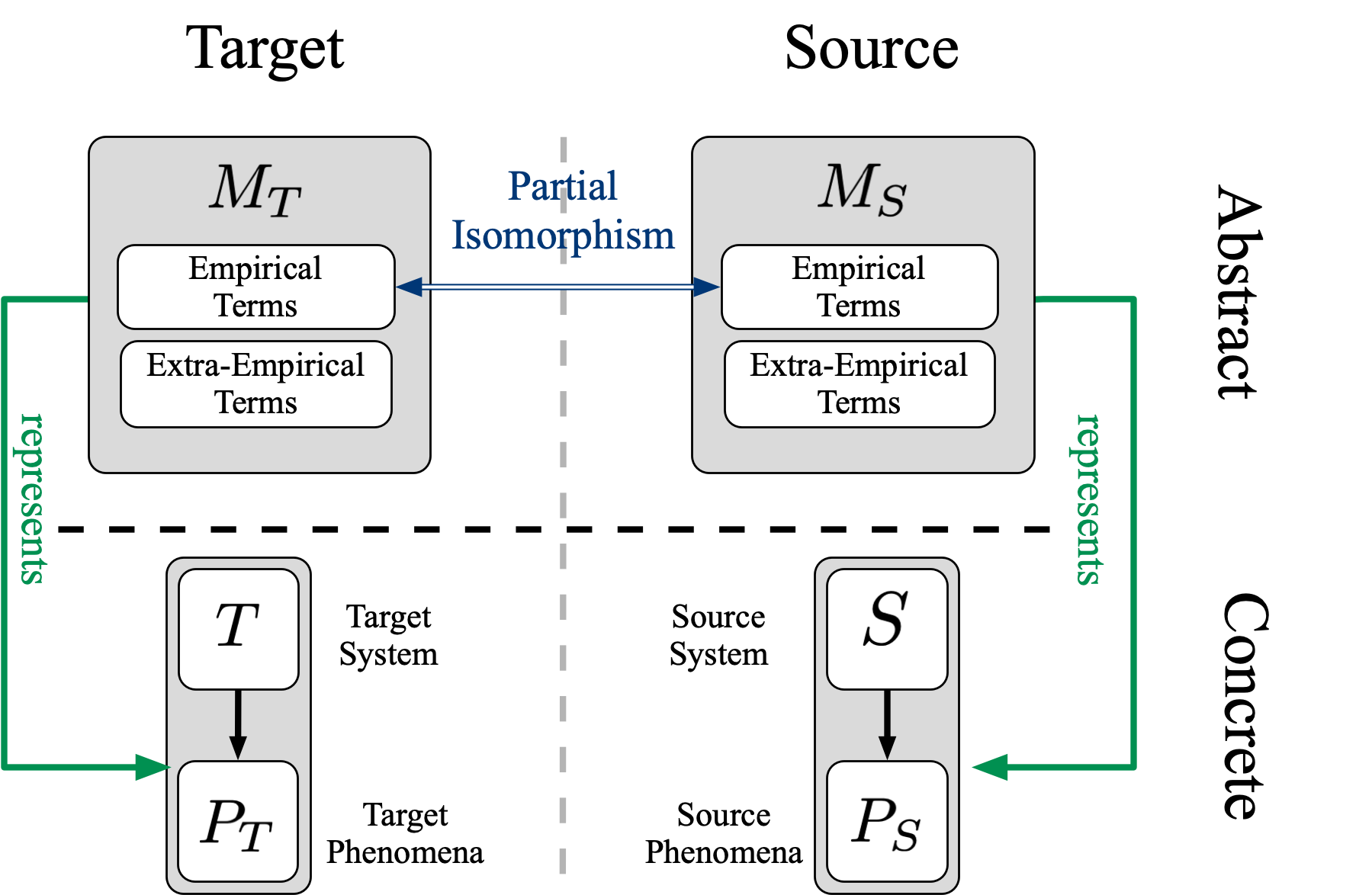}
  \caption{\textit{Analogue Simulation Schema.} The source model $M_S$ provides an empirically adequate description of a source system $S$, within some modelling domain. System $S$ displays some empirical phenomena $P_S$ that stands in a representation relation with some empirical terms in $M_S$. These empirical terms are isomorphic to some sub-set of empirical terms in the target model $M_T$. $M_T$ provides an empirically adequate description of a target system $T$, within some modelling domain. The empirical terms within $M_T$ stand in a representation relation with some target phenomena, $P_T$ which is assumed to be displayed by $T$. Analogue simulation is a scientific inference that uses the experimental realisation of $P_S$ to learn about $P_T$, based upon these formal relations (typically together with some further arguments).}\label{fig:analoguesim}
\end{figure}

With our terminology in place, we can now introduce a distinction between two types of practice in contemporary physical science, both of which might be referred to as `analogue experiments'.\footnote{We will not here consider the connection to the wide range of types of `analogue experiments' found in the context of the life sciences. Whilst there are, for example, broad conceptual connections between our analysis below and the analysis of `surrogate models' and `model organisms', the differing degree of formalisation of the two sciences render the details of physical and biological analogue experiments importantly different. See, for example, \citep{bolker:2009,levy2014,baetu:2015}. In interests of space, we will also neglect the subtle connection between analogue experiments and arguments by analogy. See \citep{Bartha2019} for further discussion.} The first, following the rational reconstruction of \citet{Dardashti2015}, we call \textit{analogue simulation}. This is when an experiment is conducted on a source system of one type of material constitution in order to learn about a target system of another type of material constitution, on the basis of a partial isomorphism that connects a sub-set of terms in the respective modelling frameworks, and which respectively provide empirically adequate descriptions of the source and target in the relevant modelling domain.\footnote{The modelling domain can be understood as a prescribed spatial, temporal, and numerical (i.e. number of atoms) scale, together with a tolerance or error margin. The isomorphism is partial in the sense that it connects only a sub-set of terms. This distinguishes analogue simulation from a duality which would be a full isomorphism between empirical terms \citep{Dardashti:2019}.} A necessary feature of this notion of analogue simulation is that the partial isomorphism must connect empirical terms defined in the two modelling frameworks. Fig.~\ref{fig:analoguesim} provides a schematic depiction of analogue simulation adapted from that provided by \citet{Dardashti2015}\footnote{Note that, whereas \citet{Dardashti2015} focus on the syntactic isomorphism between effective laws, we are focusing upon a partial isomorphism between empirical terms. So far as analogue simulation goes this is not a significance difference, we choose a different formulation only to make the comparison with analogue illustration more clear.}

As indicated by the name, analogue simulation has a lot in common with the use of computer simulation since it involves `programming' a physical source system such that there is a \textit{quantitative correspondence} between terms in the mathematical model that describes phenomena displayed by the source system and the \textit{counterpart terms} in the model that describes phenomena displayed by the target system.\footnote{This is unsurprising since core aspects of the \citet{Dardashti2015} conception of analogue simulation are drawn from earlier comparison of such practice with computer simulations due to \citet{winsberg:2010,sep-simulations-science}.} Such activity has seen a growing number of applications in research in the contemporary context of quantum simulation \citep{cirac:2012,georgescu:2013} and analogue gravity \citep{barcelo:2005,faccio:2013}.

The most vivid modern example of an analogue simulation, and the inspiration for the analysis of \citet{Dardashti2015}, is that of an analogue black hole or `dumb hole'. Here the isomorphism is between the equations describing a black hole event horizon in the semi-classical modelling domain (i.e. lengths where spacetime can be assumed to be classical) and those describing a sonic horizon in fluids in a continuum hydrodynamics modelling domain (i.e. of the order of $10^{23}$ fluid molecules). The crucial empirical terms are those describing a thermalised photonic flux, Hawking radiation, and its thermal phononic counterpart \citep{unruh:1981}. The crucial characteristic of an analogue simulation, for our purposes, is that: (i) the source--target partial isomorphism relation is a mathematical one at the level of models; and (ii) the partial isomorphism is principally established in order to exploit structural similarities that relate to \textit{empirical terms} in the respective models.

\begin{figure}
  \centering
  \includegraphics[height=0.3\textheight]{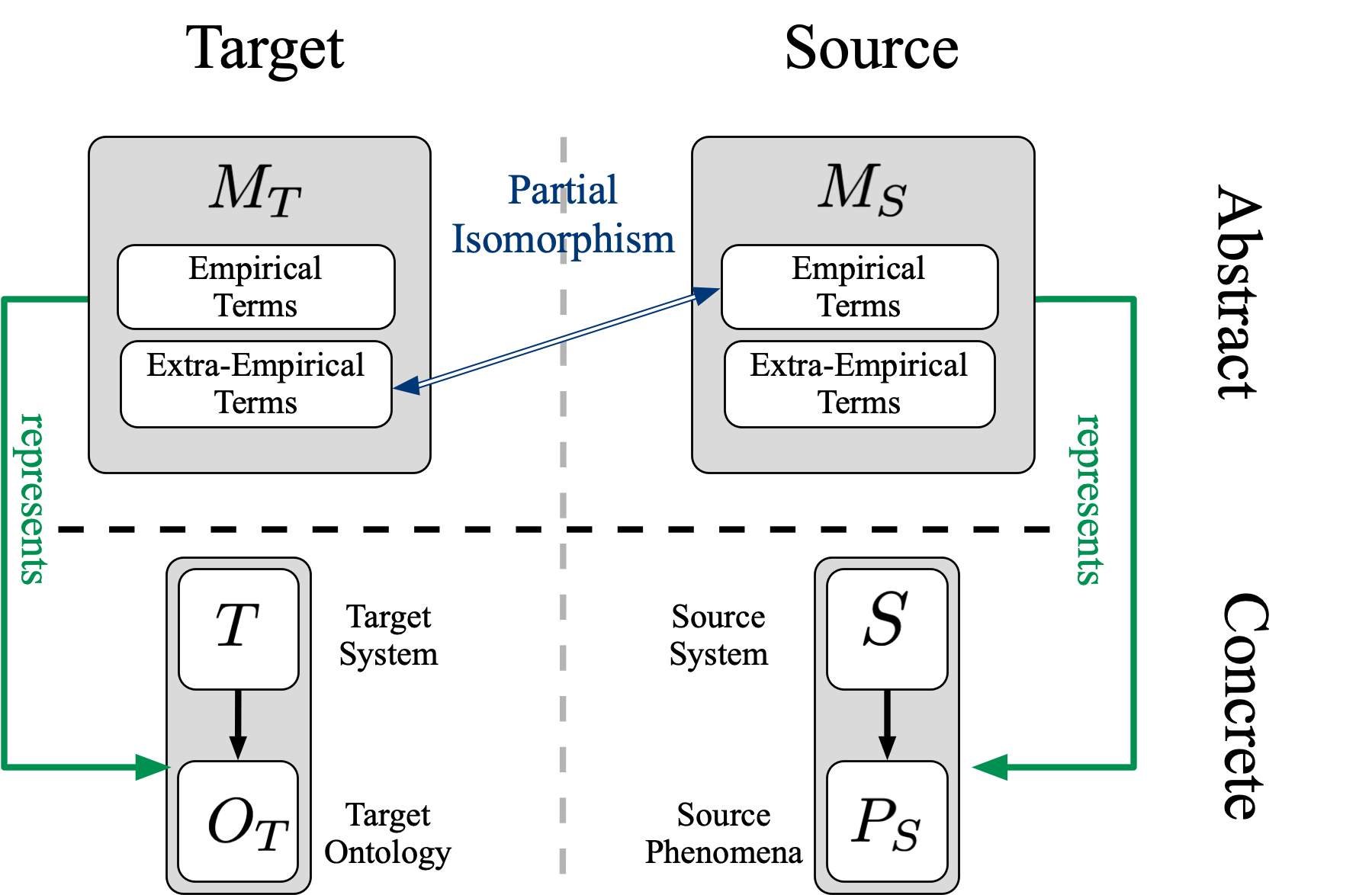}
  \caption{\textit{Analogue Illustration Schema.} The source model $M_S$ provides an empirically adequate description of a source system $S$, within some modelling domain. System $S$ displays some empirical phenomena $P_S$ that stands in a representation relation with some empirical terms in $M_S$. These empirical terms are isomorphic to some sub-set of extra-empirical terms in the target model $M_T$. $M_T$ provides an empirically adequate description of a target system $T$, within some modelling domain. The extra-empirical terms within $M_T$ stand in a representation relation with some target ontology $O_T$ which the model leads us to associate with $T$. Analogue illustration is a scientific inference that uses the experimental realisation of $P_S$ to learn about $O_T$, based upon these formal relations (typically together with some further arguments).}\label{fig:analogueill}
\end{figure}

The second type of experimental scientific practice, which can also be placed under the heading of analogue experimentation, we call \textit{analogue illustration}. This is also when an experiment is conducted on a source system of one type of material constitution in order to learn about a target system of another type of material constitution, based upon a partial isomorphism between terms in the respective modelling frameworks that provide empirically adequate descriptions of the source and target in the relevant domain. However, in this case, the basis for the putative epistemic purchase of the source system on the target system is a partial isomorphism that connects a sub-set of empirical terms in the source system modelling framework and a sub-set of extra-empirical terms in the target system modelling framework.  We thus have a parallel correspondence between certain empirical phenomena in the source system and non-phenomenal `ontological' aspects that the model associates with the target system. Fig.~\ref{fig:analogueill} provides a schematic depiction of analogue illustration.

We have chosen the name analogue illustration with the older sense of illustration as `lighting up' or `illuminating' in mind, in the sense of intellectual enlightenment. The idea is then that the analogue experiment that displays the phenomena `illuminates' aspects of the (putative) ontology of the target system. A simple example of such activity can be provided by reference to a computer illustration of the ontology of Newtonian mechanics. Consider some abstract computer code that is such that when run on a computer it generates a depiction of a homogeneous and isomorphic 3D grid. There is then a partial isomorphism (isometric mapping) between the computational `model' of the grid and the extra-empirical term corresponding to Euclidean geometry, $\mathbb{E}^3$,  in a Newtonian model of mechanics. This extra-empirical term then represents the ontology of absolute space that the Newtonian model represents via $\mathbb{E}^3$. We can thus think of the original computer grid lines on the screen as a computer illustration of Newtonian space.

We consider in detail later the sense in which the walker experiments are an analogue illustration, but for the time being we hope that the reader can see intuitively how this story will go: pilot-wave-like phenomena displayed by the bouncing oil droplets act as an analogue illustration of the particular ontology attributed to quantum systems by de Broglie's double solution interpretation. In general terms, the crucial characteristic of an analogue illustration is for our purposes that: (i) the source--target partial isomorphism relation is a mathematical one at the level of models; and (ii) the partial isomorphism is principally established in order to exploit structural similarities that relate to \textit{empirical terms} in the source models and \textit{extra-empirical} terms in the target model.

It should be clear from our definitions that analogue simulation and analogue illustration are mutually exclusive in the sense that an individual \emph{scientific inference} is either one or the other but not both since the partial isomorphism in question should either relate empirical terms in the source model to empirical terms in the target model or (exclusively) extra-empirical terms in the target model. As we shall argue below, this formal difference leads to a considerable inferential difference. We should, however, point out that this difference does not mean that a single \emph{analogue experiment} may not be used as both an analogue simulation and an analogue illustration. That is, there may exist one partial isomorphism between one sub-set of empirical terms in the source model and empirical terms in the target model and another partial isomorphism between a (possibly different) sub-set of empirical terms in the source model and extra-empirical terms in the target model. Thus, when we differentiate the two it should strictly be at the level of scientific inference rather than the experiments themselves.

\subsection{Validation}

Conventional experiments only gain their value as epistemic tools via a process of validation \citep{sep-physics-experiment}. In conventional experiments, in order to make inferences about a target system based on an experiment on a source system, a scientist requires arguments that they genuinely did learn about, first, the source system (the experiment is internally valid) and, second, that this knowledge is genuinely relevant to the features of a system or class of systems which they did not manipulate (the experiment is externally valid). In a conventional experiment, the experimenters will usually leverage the uniformity of particular material kinds as at least part of their justificatory story regarding why the source can stand in for the target in the appropriate sense.\footnote{The form of such inferences is explored in more detail in \cite{Evans:2020}.} Analogue simulations can be understood as subject to a structurally similar validation process. That is, their epistemic function requires that they must be supplemented with further arguments for internal and external validity. However, so far as external validation goes, such arguments are going to be of a very different form to those found in conventional experiment since the source and target are different kinds of material by construction. If analogue simulations are genuinely probative of empirical features of the target system it will typically only be due to these features being suitably independent of the material constitution. Building upon the work of \citet{Dardashti2015}, \citet{Thebault:2019} argues that we can think of universality arguments as playing the role of external validation in the case of the analogue black hole example of analogue simulation. \citet{hangleiter:2017} give a parallel analysis for the case of analogue quantum `emulations', and consider possible means of validation for an analogue experiment wherein a photonic source system is manipulated with the goal of learning about the phenomena of environment-assisted quantum transport in certain photosynthetic complexes.

This is where the case of analogue experiments understood as analogue illustrations looks remarkably different. When the source--target model relation is between empirical and extra-empirical terms, considerations of material similarity or material independence are not of great relevance. What is important is the isomorphism between extra-empirical terms in the target model that are difficult to conceptualise (say, the interdependence between the particle and pilot wave in the double solution ontology) and empirical terms in the source model that are visually identifiable (say, the interdependence between the walker and the surface wave). So long as that correspondence holds, the analogue experiment is usefully performing its illustrative function.

As noted above, a single analogue experiment may be used as both an analogue simulation and an analogue illustration. Consider again then the case of an analogue black hole. In this case, as well as there being an isomorphism between observables in the fluid and black hole model (e.g. Hawking radiation) we can also think of the flow of the fluid as an analogue illustration of spacetime. Crucially, the question of whether the source system adequately plays the first, `material surrogacy' role is independent from the question of whether it plays the second, illustrative role.\footnote{It is worth noting that, in this case, the experimenters are not, as it happens, particularly interested in the second function of their experiment since their main focus is on simulating empirical phenomena, and in justifying their inferences based upon universality arguments. Most likely this is because the mathematical models of a black hole in terms of Schwarzschild geometry already has a variety of simple visual illustrations (e.g. topographic diagram or Penrose diagram).} Whilst a single analogue experiment may well be simultaneously interpretable as either an analogue simulation or analogue illustration, for justificatory purposes we must be sensitive to important differences between these two functions.

\subsection{Confirmation and explanation}
\label{subsec:howpossibly}

With these distinctions and definitions in hand we can finally consider the question of what we can actually learn via analogue experiments. Again, it is worth making reference to the arguments for confirmation via analogue simulation made in the literature. What is claimed, by \citet{Dardashti2015} (see also \citet{Thebault:2019}) is that there can be confirmation via analogue simulation in circumstances where the analogue experiment is externally validated by reference to universality arguments that show that the empirical phenomena being simulated are suitably independent of the differences in material constitution between source and target system. The idea is that the universality arguments provide a common empirically grounded reason for believing in the empirical validity of both the source and target modelling frameworks within their domain of application. In virtue of that connection, inductive support for the one is then taken to be also inductive support for the other. The claims of \citet{Dardashti2015} have drawn contrasting reactions. Subsequent analysis has included extensions in terms of formal frameworks for confirmation theory \citep{Dardashti:2019,Gebharter:2020}, further exploration of the connection to conventional experiments and computer simulations \citep{Boge2018} and a (contentious) discussion of `circularity' inherent in such patterns of inference \citep{Crowther2019,Bartha2019,Evans:2020}.

Whilst the issue of confirmation in the context of analogue simulation is a live one, it is much harder to see how one could formulate an argument for conformation via analogue illustration. A little consideration of the nature of inductive evidence and the fault lines of the contemporary debate regarding scientific realism demonstrates why. Consider the inferences that a canonical empiricist and realist seek to make based upon a valid experiment. Both empiricist and realist would agree that valid experiments will, at least in some cases, allow us to provide inductive evidence in favour of the \textit{empirical adequacy} of a theoretical model; that is, the truth, within a certain domain, of empirical statements which can be derived from the model. This confirmation is built upon a correspondence between empirical terms in the model and observations drawn from the experiment.

We can set out the structure of the confirmation inference in simple Bayesian terms as follows. We denote by $H$ the claim that a particular (plausible) theory is empirically adequate in a given domain, and $E$ that a particular empirical prediction of the theory within that domain obtains. We thus have that $H \rightarrow E$, which means (almost trivially) that $P(H|E) > P(H)$, given that both $0 < P(H) < 1$ and $0 < P(E) < 1$, and so we have confirmation of $H$ by $E$. It is instructive to break this very standard inference down a little. Bayes' theorem states:
\begin{equation}
P(H|E) = \frac{P(E|H)P(H)}{P(E)},
\end{equation}
where we take the marginals, $P(H)$ and $P(E)$, to be strictly in the interval $[0,1]$, but otherwise rationally unconstrained. Given that we have assumed that $H \rightarrow E$, it follows necessarily that $P(E|H) = 1$. It then immediately follows that the confirmation measure $\triangle = P(H|E) - P(H)$ is greater than zero, and we have Bayesian confirmation of $H$ by $E$.

So far this is relatively uncontroversial. What is a subject of enduring controversy is the further move that the realist seeks to make. That is, to argue that, in at least some cases, we are licensed to make a further meta-inference to the truth (or approximate truth) of extra-empirical terms based upon (for example) inference to the best explanation or the avoidance of `miracles'. The details of this controversy are not important for our purposes. Rather, what is important is that in arguing that the extra-empirical terms can be confirmed by inductive evidence the realist should not attempt to proceed directly via induction based on the first-order evidence. This is for good reason: we have sufficient rationale to believe that theories with false extra-empirical terms can be empirically adequate in a variety of domains. The connection between the extra-empirical terms and the empirical observations is thus underdetermined to such a degree that it would be rather foolish to try to argue directly from empirical adequacy to truth in general terms. Moreover, we have, in fact, various general reasons, drawn from internal inconsistency (e.g. the well known problems with infinities in general relativity and quantum field theories), conflict between existing theories (such as quantum field theory and general relativity), and, most infamously, the history of science \citep{laudan:1981,lyons:2002,vickers:2013}, to expect that none of our current theories can be taken to be true \emph{simpliciter}.  Thus, strictly speaking, if we were going to put forward a Bayesian analysis where $H$ is now the claim that a theory is true \emph{simpliciter}, whilst we \textit{would} have that $H \rightarrow E$, and thus that $P(E | H)=1$, confirmation is blocked since we have good reason to set $P(H)=0$. Furthermore, if we weaken the realist stance such that $H$ is the claim that a theory is approximately true, not only is it no longer the case that $H \rightarrow E$, but we might plausibly set $P(E | H)$ close to zero \citep{lyons:2003}. We thus see that such direct arguments for confirmation of truth are on rather shaky ground. Realist arguments for confirmation must be founded, if they can be founded at all, on second-order evidence such as the continued success of science in general \citep{dawid:2018}.

The argument for confirmation via analogue illustration also encounters problems. In particular, given the correspondence between the extra-empirical terms in the target model and the empirical terms in the source model is engineered such that it obtains, the correspondence alone does not appear a plausible basis for an argument for confirmation. Again, a sketch of the relevant relationships in Bayesian terms will prove worthwhile. Take $H$ to be the claim that a particular extra-empirical term in our target model is true (i.e. corresponds to an `element of reality'), and take $E$ to be the successful analogue illustration of the empirical counterpart to that term in an experiment on a source system. The correspondence between extra-empirical terms in $M_T$ and empirical terms in $M_S$ is arguably independent of their respective empirical adequacy or truth since it has been engineered to exist. Thus, it seems difficult to avoid the conclusion that we should set $P(E | H)=P(E)$, and thus $P(H| E)=P(H)$, which implies $\triangle=0$, and thus that no confirmation can obtain. In this context, any argument for confirmation would need to be based upon something beyond the relevant correspondence obtaining. That said, we do not rule out the possibility that such an argument could be given in principle. It might perhaps be based upon the combination of inference to the best explanation and the `surprising nature' of the mere possibility of establishing correspondence. However, we do not see any obvious means of formalising such an argument. Moreover, since it seems perfectly possible for us to construct analogue illustrations of false theories this kind of explanationist strategy does not appear a very fruitful one.

This brings us to the question of what analogue illustrations are for, if not confirmation. A tentative answer, that we will put to use below, comes from the idea of \textit{how-possibly explanation}. First introduced by \citet[\S VI]{Dray57}, how-possibly explanation is a rival kind of explanation to deductive-nomological explanations.\footnote{For more discussion on how-possibly explanation see \citep{Forber10,Bokulich14,Cuffaro15}. \citet{hangleiter:2017} argue that how-possibly understanding can be understood as a supplementary function of certain forms of analogue simulation.} \citet{Persson12} provides a productive way, for our current purposes, to conceive of how-possibly explanation: how-possibly explanation offers a \emph{potential} how-explanation that explores the space of possible explanantia for some explanandum. As part of formulating such how-possibly explanation, we must metaphysically and epistemically `bracket', in the sense that the explanation is not about how the world actually is, nor do we know whether the explanation is true. \citet[p.282]{Persson12} claims that the context of this type of how-possibly explanation
\begin{quote}
  is typically one of discovery, hypothesis generation, or the exploration of a range of possible explanations in a research environment where the explanandum phenomenon is accepted as a fact and now needs to be integrated with the system.
\end{quote}
\citet{Reutlinger17} rationally reconstruct the epistemic goal of toy modelling in science, claim that how-possibly explanation is one such goal, and go on to identify at least three epistemic functions of how-possibly explanation that endow it with value for the scientist. The first is a modal function: how-possibly explanation is valuable when ``scientists want to understand whether and why some phenomenon is possibly or necessarily the case'' \citep[p.26]{Reutlinger17}. The second epistemic function is a heuristic function: how-possibly explanation is valuable when it stimulates further investigation into a more accurate model of the target system. The final epistemic function is a pedagogical function: how-possibly explanation is valuable when used for illustrative purposes to ``enable students and researchers to quickly grasp the idea behind\ldots the description of a phenomenon'' \citep[p.26]{Reutlinger17}.

We think it is plausible to take analogue illustration to be a fruitful means to provide all three of these functions. The following section will articulate why, based upon the example of the bouncing oil droplet and pilot wave theory.

\section{What can we learn from the walker experiments?}
\label{sec:Parisanalogue}

In this section we first introduce a reason to be optimistic about the possibility of the walker experiments being an analogue simulation of quantum phenomena (\S\ref{subsec:borghesi}). We then provide a more significant reason to treat them as analogue illustrations (\S\ref{subsec:taget}). In doing so, we argue that the walker experiments cannot provide inductive evidence for a pilot wave ontology for quantum phenomena, whether de Broglie's double solution theory or not. Finally, in \S\ref{sec:how}, we consider the prospect that the walker experiments might still be valuable in terms of the how-possible explanation they provide.

\subsection{Simulation and confirmation}
\label{subsec:borghesi}

An analogue simulation requires a partial isomorphism between empirical terms in two modelling frameworks that we take to be adequate models of relevant source and target systems. It is clear in the current context that the target modelling framework is de Broglie's double solution theory, as a model of quantum phenomena. We then need to identify a model of the fluid mechanical walker system. Against this background, \citet{Borghesi17} has developed a classical toy model in light of the fluid mechanical walker system and de Broglie's double solution theory.

Motivated by the desire to model the walker system in a relativistically covariant manner,\footnote{Since it is a virtue of any modelling framework to have as large a domain of validity as possible, the relativistic covariance of a model of the walker system is inherently desirable.} Borghesi's model consists of a vibrating elastic medium carrying a transverse wave, $\phi$,\footnote{$\phi$ denotes a transverse displacement and not a phase.} and a point-like `concretion' (a very concentrated heterogeneity of the medium itself). The mechanism of the model is comprised of an equation of motion for the concretion, in which the motion of the concretion is deflected by the gradient of $\phi$ at its location, and a wave equation that can be interpreted as an inhomogeneous Klein-Gordon-like equation in the presence of a wave source localised at the position of the concretion and dependent upon its vibration \citep[p.938]{Borghesi17}. A curious simplification of this toy model leads to interesting consequences.

To reflect better the interdependence in the walker experiments between the walker and the surface wave, Borghesi assumes that the concretion no longer acts as the wave source after ``a kind of self-adaptive phenomenon between the transverse wave $\phi$ and the concretion'' \citep[p.939]{Borghesi17}. This establishes an `intimate harmony' between the wave and the concretion, which Borghesi labels `symbiosis', such that there is no back-reaction of the concretion on $\phi$ (as is the case for the de Broglie guiding equation \citep{Holland05}). The regime in which symbiosis is possible requires that in the proper reference frame of the concretion, the concretion is located at a local extremum of $\phi$. In this regime, the mass continuity equation leads to a concretion speed constant in time, and so reflects conservation of particle energy in the absence of an external potential, and it also leads to a relation that Borghesi calls the $\phi$-guidance formula. In the low-velocity approximation, the $\phi$-guidance formula reduces to be syntactically isomorphic to de Broglie's guiding equation.

Perhaps more remarkably, when the concretion is not the source of the wave, and assuming ``that the time period of transverse oscillations is much shorter [than] the characteristic evolution time of the (perceivable) motion of the concretion'', then the motion of the concretion (in symbiosis with the wave) is guided exclusively by the $\phi$-guidance formula \citep[p.941]{Borghesi17}. That is, averaged over one transverse oscillation, the equation of motion for the concretion does not play any role in its motion. This `cancellation' of the equation of motion in the symbiotic regime, due to the $\phi$-guidance formula, ensures that the velocity of the concretion is directly proportional to the gradient of the wave, in contrast to an equation of motion which relates an external potential to particle acceleration.

In the symbiotic regime, where the concretion satisfies the $\phi$-guidance formula, the following can also be derived from the model: a `wave potential', syntactically isomorphic to the quantum potential derived by \citet{deBroglie27a} and \citet{Bohm}; an expression for the energy of the concretion in the low-velocity limit that contains a `rest mass' term and an additional energy term, $E_{c}$, equal to a coefficient multiplied by an additional pulsation beyond the transverse vibrations of the wave, $\omega$; a proportionality coefficient between wave and particle characteristics of the model that, when represented by the term $\hbar_{\text{exp}}$, renders the additional energy term $E_{c} = \hbar_{\text{exp}}\omega$, isomorphic to the Planck-Einstein relation;\footnote{It should be noted that $\hbar_{\text{exp}}$ here is not Planck's constant, but is rather a ``proportional coefficient between wave characteristics and particle characteristics'' of the concretion model \citep[p.945]{Borghesi17}. It appears that this specific notation was chosen to emphasise the isomorphism with the Planck-Einstein, and other quantum, relations.} and a low-velocity approximation of the Klein-Gordon-like wave equation without source that is isomorphic to the free Schr\"{o}dinger equation. Moreover, these expressions can be combined to produce an expression for the energy and momentum of the concretion that are isomorphic to expressions that represent the energy and momentum of a quantum system. On this final isomorphism, \citet[p.946]{Borghesi17} points out: ``This point confirms in our system what de Broglie had suggested, here restricted to energy and momentum: the particle accounts for quantities commonly attributed to the wave-like nature of the system in quantum mechanics''.

Based on these claims, we can see that there is a partial isomorphism between Borghesi's fluid dynamical modelling framework for the walker system and de Broglie's formulation of pilot wave theory. Moreover, and this is slightly more contentious, in so far as de Broglie toyed with some form of the Klein-Gordon equation as the wave equation for his $u$-waves, Borghesi derives a wave equation (from a principle of least action) that can be interpreted as an inhomogeneous Klein-Gordon-like equation in the presence of a wave source (although Borghesi himself does not go so far as to endorse any analogy on this point). It would be incorrect to claim that there is an isomorphism at play between these latter two wave equations, but the two wave equations have similar form (and, recall, de Broglie never found the precise wave equation for his $u$-waves).

However, on closer inspection we see that the isomorphism is not between empirical terms of the two modelling frameworks. The partial isomorphism here is between the structural equations describing the concretion and elastic medium in Borghesi's classical toy model and the $u_{i}$ and their associated singularities in de Broglie's pilot wave theory. As such, this partial isomorphism establishes a correspondence between empirical terms in Borghesi's model, such as the energy, momentum, and position of the concretion, the slope of the transverse displacement of the elastic medium, and a modulating wave associated with the presence of the concretion (representing properties of the walker, surface wave, and probability wave, respectively), and key \textit{extra-empirical} terms in de Broglie's pilot wave theory, including the quantum phase and the pilot wave itself.\footnote{One could argue that the energy, momentum, and position of each quantum in a pilot wave theory also count as extra-empirical terms, at least in the regime where evolution is unitary. This is because, according to pilot wave theory, the precise energy, momenta, and position of the quanta comprise `hidden' variables. (This is, of course, not to mention the constraints on precise values of momentum and position supplied by the uncertainty principle.)} Establishing a partial isomorphism between important terms is indeed required to argue that the walker experiments might count as an analogue simulation of pilot wave theory, but the partial isomorphism required must correlate empirical terms from the analogous modelling frameworks. As this partial isomorphism does not do this, we claim that this is not an example of analogue simulation.

Furthermore, even if the right relations could be established, it is by no means clear that an argument for confirmation of pilot wave theory via analogue simulation could be run. First, let us consider again the structure of the arguments in the literature for confirmation via analogue simulation in the context of analogue black hole experiments \citep{Dardashti2015,Thebault:2019,Dardashti:2019}. All such arguments rely upon the analogue experiment in question being externally validated by reference to universality arguments that show that the empirical phenomena being simulated are suitably independent of the differences in material constitution between source and target system. The idea is that the universality arguments provide a common empirically grounded reason for believing in the empirical validity of both the source and target modelling frameworks within their domain of application. In virtue of that connection, inductive support for the one is then taken to be also inductive support for the other. There is no obvious possibility for such a universality argument to be provided in this case. Second, and more significantly, what would be confirmed if such confirmation could be argued for would be the empirical adequacy of the pilot wave theory, not its truth. And we already know that the theory is empirically adequate! As such, considered as an analogue simulation we could not, even in the best of circumstances, be said to have learnt much from the walker experiments. However, we need not take the function of the walker experiments as an analogue simulation, as we have good reason to understand the principal function of the experiments as an analogue illustration of the pilot wave ontology.

\subsection{Illustration and confirmation}
\label{subsec:taget}

\begin{figure}
  \centering
  \includegraphics[height=0.3\textheight]{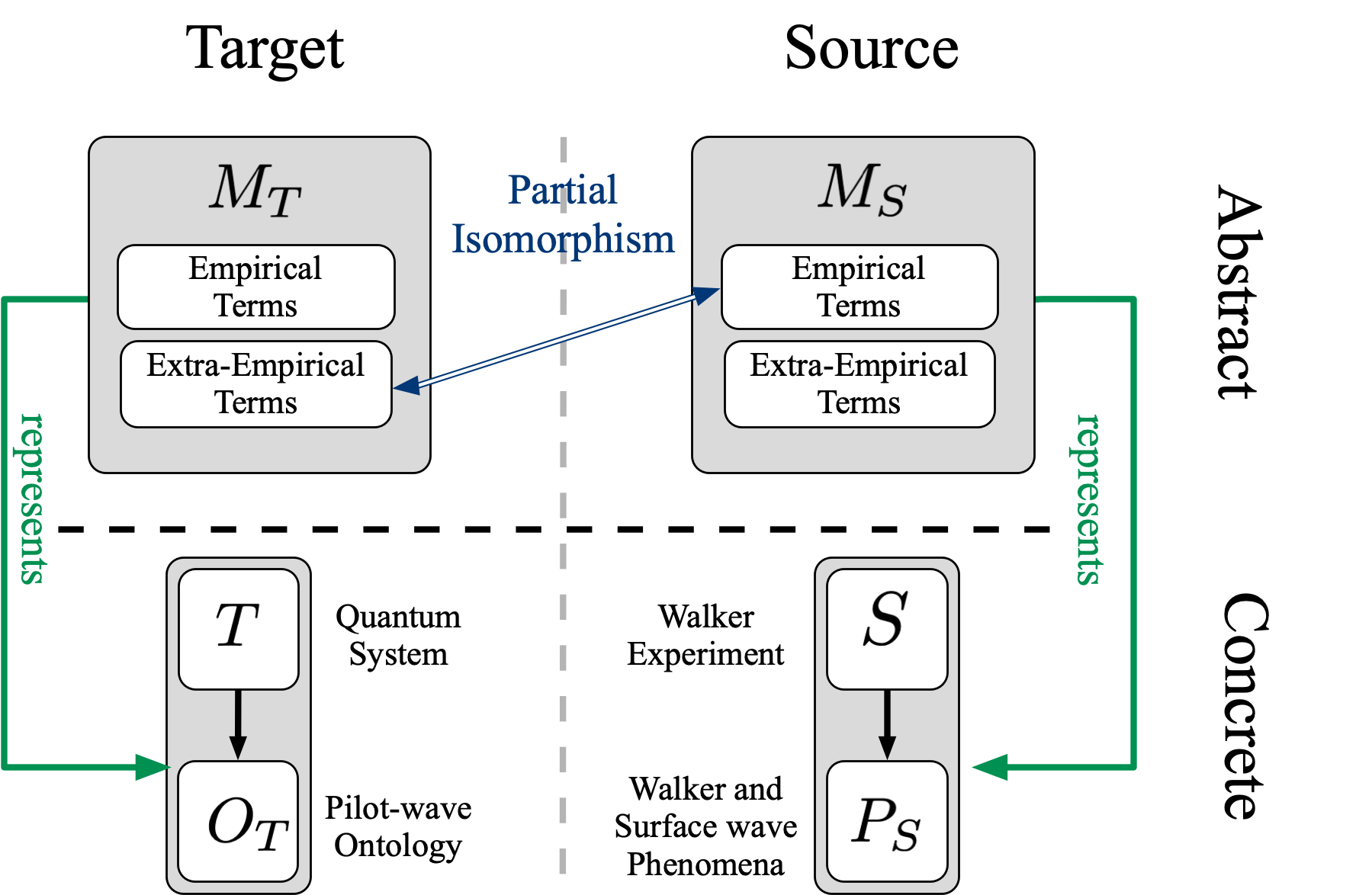}
  \caption{\textit{Analogue Illustration Walker Experiments.} See description in text.}\label{fig:walker}
\end{figure}

Let us reconsider our schema for analogue illustration in the specific context of the walker experiments and pilot wave theory (see Fig.~\ref{fig:walker}). On the `source' side of the diagram the transcription is straightforward. The source system, $S$, is the fluid mechanical walker system. The phenomena, $P_{S}$, are the range of typically quantum behaviours displayed by the walker system such as diffraction, interference, quantised orbits, and tunnelling. We can collect them together as walker and surface wave phenomena. The modelling framework, $M_{S}$, is Borghesi's classical toy model of the symbiotic concretion and elastic medium. On the `target' side of the diagram the modelling framework, $M_{T}$, is de Broglie's pilot wave theory. The target system is a quantum system and the target ontology, $O_{T}$, is that of the pilot wave theory.

The walker experiments are providing us with a macroscopic illustration of wave-particle duality as realised in the ontology of pilot wave theory based upon a partial isomorphism between empirical terms in Borghesi's classical toy model and extra-empirical terms in de Broglie's formulation of pilot wave theory. What inferential purpose do scientists have creating such an analogue illustration? On account of the argument in \S\ref{subsec:howpossibly}, we should not think that inductive support for the ontology of the target system is provided by analogue illustrations. This seems true for the current case in particular. Quantum mechanics, as a modelling framework, notoriously underdetermines quantum ontology, and as striking as the typically quantum behaviour of the walker experiments is, they do not seem to provide epistemic warrant in favour of a pilot wave ontology for quantum mechanics.

Given that it is difficult to suggest that the walker experiments can be providing us with inductive evidence regarding de Broglie's interpretation of quantum mechanics, or quantum mechanics more generally, one might conclude that the walker experiments are epistemically worthless. However, this seems a little too quick: after all, the reproduction of so many typically quantum phenomena, and the concrete illustration of wave-particle duality via a pilot wave, particularly of the sort envisaged in de Broglie's double solution theory, is extraordinary. We contend that the epistemic value of the walker system should be understood in terms of how-possibly explanation.

\subsection{Explanatory aspect}
\label{sec:how}

We can proffer here a rational reconstruction of the epistemic goals of the experimental teams undertaking the walker experiments in the context of the conception of how-possibly explanation from \S\ref{subsec:howpossibly}. Since any particular phenomenon only makes sense as an explanans in the context of a scientific model or theory, we take the explanans in the current case to be the walker experiments in the context of de Broglie's double solution theory. It then seems reasonable to suppose that the experimental teams undertaking the walker experiments are taking the explanandum to be the mechanism that underlies wave-particle duality in the double solution theory: by what physical mechanism, according to the double solution pilot wave theory, could the interdependence between particle and wave be established? The walker experiments, in combination with the double solution theory, demonstrate a possible (classical) mechanism through which that interdependence can be established. As an explanation of this explanandum, it does not seem that the Paris and MIT experimenters presume that the experiment is a representation of how the world actually is, nor that they know whether this mechanism for wave-particle duality is true of the double solution pilot wave theory. In other words, claims about the world have been metaphysically and epistemically bracketed. Furthermore, the walker experiments surely must be taken to be performed in the context of ``discovery'' and ``hypothesis generation'' in the face of acceptance that quantum mechanics renders wave-particle duality as a fact that is in need of integration, metaphysically speaking, with the rest of contemporary physics. Thus, given Persson's conception of how-possibly explanation, it seems a suitable characterisation of the utility of the walker experiments. But is there value in this utility?

Recall the three epistemic functions of how-possibly explanation that endow it with value according to \citet{Reutlinger17}: the modal function; the heuristic function; and the pedagogical function. All three can be seen to hold for the walker experiments: the first since the experiments are concerned with whether it is possible to represent coherently wave-particle duality by means of a pilot wave mechanism of the type envisaged by de Broglie's double solution theory; the second since the fluid mechanical model provides a concrete illustrate of the ontology of de Broglie's double solution theory, and therein a pilot wave mechanism for wave-particle duality, and this suggests that de Broglie's incomplete double solution theory might be worthy of renewed exploration; and the third since the walker experiments patently provide a clear illustrational tool for visualising a classical mechanism that underlies typically quantum behaviour, as well as more specifically a tool for visualising the pilot wave ontology of the double solution theory.\footnote{One only need search YouTube for an array of examples of the pedagogical value of the walker experiments.}

On account of these epistemic functions of how-possibly explanation, the walker experiments have definite epistemic value as an analogue illustration: they provide us with how-possibly understanding of wave-particle duality via a pilot wave ontology. In this sense, we take this instance of analogue illustration to be an example of the \textit{passive} dimension that how-possibly explanation might play in scientific inference, explored with respect to toy models by \citet{Reutlinger17}. More than this, however, we take these experiments to demonstrate a \textit{new role} for how-possibly explanation in scientific inference: the \textit{active exploration} of a target ontology via manipulation of the source phenomena, of the sort that we usually take to constitute traditional scientific inquiry. Crucially, this possibility is underpinned by the formal partial isomorphism established between the source and target in an analogue illustration. This active form of how-possibly explanation might plausibly be generalisable to other scientific inferences beyond analogue illustration and, moreover, highlights a further epistemic value of these experiments.

\section{Final thoughts}
\label{sec:final}

There is evidently something remarkable about the fact that a series of experiments consisting of a vibrating bath of silicone oil sustaining bouncing droplets on its surface can provide a concrete illustration of the ontology of a long forgotten pilot wave theory of quantum mechanics (one developed when quantum mechanics was in its infancy, to boot). As we have seen here, however, we must not be too hasty in considering the debate around the metaphysical consequences of quantum mechanics to be any closer to resolution as a result. When supplemented with suitable additional arguments, the framework of analogue simulation is capable of providing epistemic licence to raise our probability that we will observe some phenomenon of a target system on account of observing some phenomenon in a source system. However, the relation between the walker experiments and de Broglie's double solution theory does not fit naturally into the analogue simulation framework. Rather, these experiments are better understood as an example of analogue illustration. Significantly, we should not take analogue illustration to establish anything beyond mere plausibility for the existence of elements of reality corresponding to extra-empirical terms, such as the ontology of quantum mechanics claimed by the double solution theory. Thus the walker experiments provide us with no justification to say anything more about quantum mechanics beyond the plausibility of some ontological framework -- and pilot wave theory is already considered a plausible ontological framework for quantum mechanics.

This notwithstanding, the walker experiments plausibly do provide us with a how-possibly explanation for the mechanism that underlies wave-particle duality in the double solution pilot wave theory. Not only does this provide a clear pedagogical tool for visualising both a classical mechanism underlying typically quantum behaviour and the pilot wave ontology of the double solution theory, but it also suggests that further investigation in quantum foundations into de Broglie's incomplete double solution theory could plausibly prove fruitful. Furthermore, it is perhaps worth reflecting upon the remarkable fact that the walker experiments give us an example where, through analogue illustration, we can learn, in the sense of gaining how-possibly understanding, about the putative ontology of a target system via an experiment. As such, in considering the nature and goal of these experiments, we have isolated an unconventional and novel form of scientific inference that should be of considerable pertinence to philosophers interested in the nature of scientific explanation in general.

To conclude, there are many more points of interest concerning the relation between the walker experiments and quantum mechanics that have not been addressed in this paper. One such issue is the classicality of the fluid mechanical model. Attempts in recent years to describe typically quantum correlations in terms of classical causal modelling frameworks \citep{Spirtes00,Pearl09} have notoriously failed, and significant steps have been taken towards a quantum version of causal modelling \citep{LeiferSpekkens2013,CavalcantiLal14,CostaShrapnel16,Allen17}. Accordingly, one must attach a healthy level of scepticism to the idea that a classical fluid mechanical model could reproduce typically quantum correlations, as any classical mechanism looks likely to be too impoverished to do so (despite the arguments of \citet{Vervoort16}). And this is the rather sizeable elephant in the room in the context of asking what bouncing oil droplets can tells us about quantum mechanics: the walker experiments thus far performed evidently lack the capacity to act as either analogue simulations or analogue illustrations of quantum entanglement. As \citet[p.287]{Bush15} puts it, since the typically quantum behaviour of the walker arises as a function of chaotic dynamics, ``[t]he question remains open as to whether some combination of intrusive measurement and chaotic pilot wave dynamics might give rise to a hydrodynamic analog of entanglement.'' Progress on this issue could significantly change the nature of the relation between the walker experiments and quantum mechanics as we have presented it here but, as it stands, we do not expect the walker experiments to be an analogue illustration or simulation of any entanglement-based quantum phenomena that involve the violation of Bell-type inequalities.

That said, we do not believe that it is plausible to dismiss the walker experiments as scientifically valueless on the basis that they cannot as yet serve as analogue simulations or analogue illustrations of quantum entanglement. While one might portray quantum entanglement, or the measurement problem, as \emph{the} defining feature of quantum theory, another legitimate candidate for this mantle is wave-particle duality. From this perspective, in so far as the walker experiments provide a concrete illustration of wave-particle duality, and how-possibly explanation for the mechanism that underlies wave-particle duality in the double solution pilot wave theory, there remains in the walker experiments epistemic value for quantum foundations research. Alternatively, putting this idea on its head, one might also argue that the walker experiments support the complementary conclusion: since the walker experiments show that a classical fluid dynamical model can illustrate wave-particle duality, one might read that as indicating that wave-particle duality is \textit{not} a distinctively quantum phenomenon. From this one might conclude that in so far as the question of distinctly quantum phenomena goes, there is no great epistemic value to be gained for quantum foundations research by investigating wave-particle duality. In this context, the potential value of these experiments can be understood in terms of the incomplete or failed analogy that they instantiate. The abiding value of these experiments would then be that of a negative heuristic.

\vspace{5mm}
\singlespacing
{\footnotesize \noindent \textbf{Acknowledgments:} This work grew out of a Research Group Fellowship in 2015 at the Munich Center for Mathematical Philosophy, with fellow group members Radin Dardashti, Matt Farr, and Alex Reutlinger. We are grateful to the hospitality of Ludwig-Maximilians-Universit\"{a}t and Stephan Hartmann for hosting us during the early stages of this research. For valuable discussion, comments, and feedback we are greatly appreciative to Guido Bacciagaluppi, Christian Borghesi, Paul Teller, Eric Winsberg, two anonomous referees, and to audiences in Brisbane and Canberra. PWE's work on this paper was supported by the Templeton World Charity Foundation (TWCF 0064/AB38), the University of Queensland, and the Australian Government through the Australian Research Council (DE170100808). KT's work on this paper was supported by the Arts and Humanities Research Council, UK (AH/P004415/1).}

\providecommand{\noopsort}[1]{}

\end{document}